\documentclass[preprint,showpacs,prb]{revtex4}
\usepackage{graphicx}

\begin{document}

\title{The Structure and Stokes Shift of Hydrogenated Silicon
Nanoclusters}

\author{Lucas Wagner$^1$, Aaron Puzder$^2$, Andrew
Williamson$^2$, Zachary Helms$^1$, 
 Jeffrey C. Grossman$^2$, Lubos Mitas$^1$, Giulia
Galli$^2$ and Munir Nayfeh$^3$}
\affiliation {$^1$Department of Physics, North Carolina State University,
 Raleigh, NC 27695-8202\\}
\affiliation{$^2$ Lawrence Livermore National Laboratory, Livermore,
CA 94550 \\}
\affiliation{$^3$Department of Physics, University of Illinois at
Urbana-Champaign,Urbana, IL 61801\\}

\date{\today} 

\begin{abstract}
  We evaluate the optical gap and Stokes shift of several candidate
  1~nm silicon nanocrystal structures using density functional and
  quantum Monte Carlo (QMC) methods. We find that the combination of
  absorption gap calculations and Stokes shift calculations may be
  used to determine structures.  We find that although absorption gaps
  calculated within B3LYP and QMC agree for spherical, completely
  hydrogenated silicon nanocrystals, they disagree in clusters with
  different surface bonding networks. The nature of the Stokes shift
  of the ultrabright luminescence is examined by comparing possible
  relaxation mechanisms.  We find that the exciton which reproduces
  the experimental value of the Stokes shift is most likely a state
  formed by a collective structural relaxation distributed over the
  entire cluster.
\end{abstract}


\maketitle

\section{Introduction}
  
The investigation of semiconducting nanoclusters is one of the most
promising directions in the search for new materials to construct
optical and electronic devices, new laser materials, and biological
markers~\cite{yoffe,chen,chan}.~ The properties of nanosize clusters
are typically very different from their parent bulk compounds; for
example, surface passivated silicon nanoclusters show a number of
interesting effects such as ultrabright
luminescence~\cite{furukawa,wilson,schuppler,vanbuuren,holmes,belomoin2002}
and nonlinear optical effects~\cite{nayfeh2001}, whereas crystalline
bulk silicon is optically uninteresting because of its small and
indirect gap.  Previous theoretical studies have employed various
methods to model and understand the basic structural and electronic
properties of nanosize clusters such as tight-binding~\cite{proot92},
empirical pseudopotentials~\cite{wang94}, density functional theory
(DFT)~\cite{delley,ogut}, GW-Bethe Salpeter
(GW-BSE)~\cite{rohlfing98,benedict}, and quantum Monte Carlo
(QMC)~\cite{puzder2002a}.~ Despite the significant progress that has
been made, especially in interpreting the properties of larger
clusters ($>$ 2 nm , $>$ 500 atoms), many open questions still exist
as to how new physical effects begin to dominate the electronic
structure of these clusters as the surface to volume ratio increases.
These effects include the possibility of surface states,
reconstructions, and impurities.

Recent optical measurements of silicon nanocrystals synthesized in
macroscopic quantities using a process of first etching and then
sonically breaking up a silicon wafer~\cite{belomoin2000} appear to
have predominantly spherical shapes, show ultrabright luminescence,
and are produced in remarkably uniform batches.  The smallest silicon
nanoclusters in the range of 1 to 1.2 nm~\cite{belomoin2000}
synthesized through this method exhibit significantly different
properties compared with predictions from previous studies such as
smaller absorption gaps and different red shifted emission peaks
(i.e.~the Stokes
shift)~\cite{proot92,wang94,vasiliev2001,williamson2002,martin,takagahara,allan96}.~
Understanding the differences in these two characteristics call for
high quality electronic structure calculations which can precisely
describe the effects of electron correlation in ground and excited
states, charge transfers, and different surface bonding networks.
 
In this article, we present a state-of-the-art computational study of
$\sim$ 1 nm hydrogenated Si nanocluster prototype systems in order to
understand the interplay between their electronic and optical
properties, their surface states and structures, and to interpret
their absorption and emission processes.  We employ a combination of
high accuracy {\em ab initio} methods including DFT (using the local
density approximation (LDA) and the Perdew Burke Ernzerhof
(PBE)~\cite{perdew} and B3LYP flavors of the generalized gradient
approximation (GGA)) and QMC to elucidate the electronic and
structural properties of the most relevant structural prototypes.  We
focus on systems with 50 to 70 total atoms which are consistent with
the observed sizes that emit in the blue or at the UV edge ($\sim$
2.90 eV).~ We find that surface dimerization is considered as the best
possible mechanism for the occurrence of smaller gaps and Stokes
shifts than would otherwise be expected in this size
range~\cite{mitas2001}.~ We use our derived models to study the
structural and electronic effects related to the observed Stokes
shift, namely the energy difference between absorbed and emitted
photons and the character of the corresponding exciton state.  The
high accuracy of our QMC calculations enables us to deduce that the
mechanism most likely to be responsible for the Stokes shift in these
systems is due to {\em a global relaxation of the cluster} rather than
the stretch mode of a single surface dimer~\cite{allan96}.~ Our
results demonstrate that the accurate calculation of a combination of
properties is critical for a full understanding of surface
reconstructions, doping, and the Stokes shift of nanoclusters in this
size range.  Finally, we find that despite its agreement in
determining optical gaps for spherical silicon nanocrystals, B3LYP
compares less favorably with QMC results for systems with
reconstructed surfaces or with oxygenated surfaces.
 
\section{Calculational Methods}
 
\begin{figure}[ht]
\includegraphics*[width=\columnwidth]{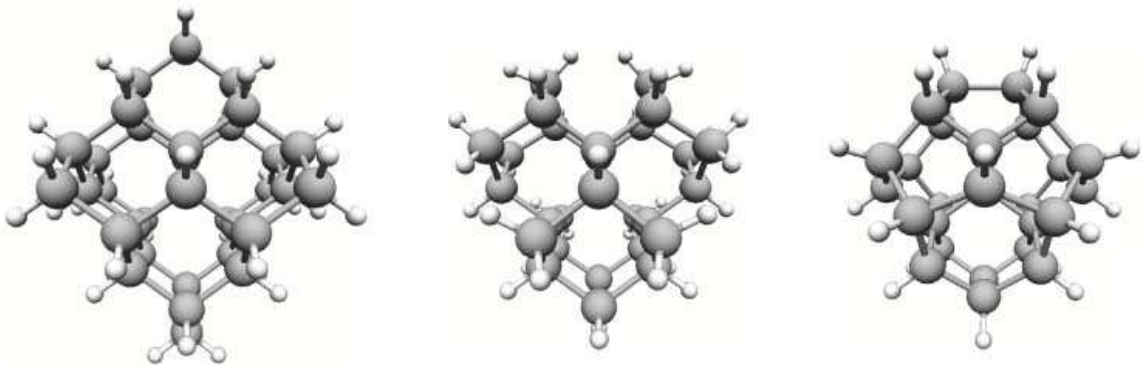}
\caption{ Atomic structures of Si$_{35}$H$_{36}$, Si$_{29}$H$_{36}$, 
  and Si$_{29}$H$_{24}$ nanocrystals. The larger spheres represent Si
  atoms while the smaller ones the hydrogens. Note the surface
  reconstruction of the top dimer from Si$_{29}$H$_{36}$ to
  Si$_{29}$H$_{24}$.  }
\label{fig:sih}
\end{figure}

To construct structural prototypes, we start from the bulk silicon
lattice and choose a spherical region as this has the smallest surface
area for a given number of atoms.  The dangling bonds on the surface
are saturated with hydrogen and the terminated Si atoms are classified
into -SiH, -SiH$_2$ and -SiH$_3$ types.  The structures with -SiH$_3$
are discarded as they are energetically less favorable and therefore
prone to reactions with the environment.  Considering prototypes with
-SiH and -SiH$_2$ terminations in the 1-1.2 nm size regime leads to
two likely ideal structures, Si$_{29}$H$_{36}$ and Si$_{35}$H$_{36}$
(Fig.~\ref{fig:sih}).~ In some cases, DFT calculations using a
planewave basis and norm-conserving pseudopotentials~\cite{jeep} are
used to establish the equilibrium geometries and the electronic
structure of these systems with a kinetic energy cutoff of 35~Ry, a
Gianozzi pseudopotential for hydrogen~\cite{gianozzi}, and a Hamann
pseudopotential for all other atoms~\cite{hamann}.~ In other cases,
Gaussian all-electron calculations using a 6-311G$^*$ basis is
used~\cite{gauss}.~ As we have previously observed~\cite{puzder2002a},
the relaxed structures are very similar when calculated using either
LDA or gradient corrected (PBE) functionals and are fairly close to
the original bulk derived structures with small adjustments of the
bond lengths and angles, predominantly of the surface atoms.  For each
cluster, the highest occupied molecular orbital (HOMO)/ lowest
unoccupied molecular orbital (LUMO) gap is calculated using LDA, PBE,
and the B3LYP functional.

Although DFT techniques produce accurate minimum energy structures of
silicon nanoclusters and also reliably predict the trends in the
optical gap of a given structural type as a function of
size~\cite{puzder2002a}, a more accurate approach, for example, QMC
which takes many-body effects into account is required to predict the
difference in optical gaps between different classes of structural
prototypes, such as clusters with different surface passivants,
clusters with reconstructed surfaces, or clusters with amorphous-like
geometries.  In addition to the computed DFT gaps, we therefore adopt
a previously described QMC procedure~\cite{puzder2002b,williamson2001}
to perform QMC calculations using the CASINO code~\cite{casino} for
the optical gaps discussed in this paper.

\section{Predicted Structure Through Comparison with Absorption Gap}
  
The smallest experimentally measured clusters are determined to be
$\sim~1$~nm in diameter~\cite{belomoin2000}.~ 1~nm clusters are too
small to observe crystallinity using transmission electron microscopy
(TEM), so the structure must be discerned from other experimental
methods, or deduced from physical properties such as the absorption
gap and the Stokes shift.  Indeed, even if TEM or other methods were
capable of resolving the crystallinity of 1~nm clusters, it could not
predict the surface chemistry.  In a previous work, we compared the
absorption gap of 1~nm spherical silicon nanoclusters with two types
of reconstructed surfaces~\cite{mitas2001}.~ Amongst the various
structures analyzed, only clusters with reconstructed dimers yielded
results consistent with the experimentally measured gaps.~ However, in
addition to all possible oxygenated clusters which were not
considered, a recent molecular dynamics calculation predicted that
amorphous-like 1~nm clusters can have similar gaps to those with
reconstructed dimers~\cite{draeger}.~ Here, we have recalculated the
gaps of these ideal and reconstructed clusters, now using linear
scaling QMC~\cite{williamson2001,casino} which has allowed us to
obtain much smaller statistical errors, and compared them with the
double cored clusters~\cite{draeger} as well as various oxygenated
clusters.~ For comparison, we also compute all relevant HOMO/LUMO gaps
within DFT using various functionals, focusing on the B3LYP functional
in order to make a more complete comparison of this functional with
computationally demanding many-body QMC results.

Table~1 shows that the optical absorption gaps as calculated within
QMC of two completely hydrogenated spherical crystalline nanoclusters
about 1~nm in diameter, Si$_{29}$H$_{36}$ and Si$_{35}$H$_{36}$ (5.3
and 4.9 eV) are larger than our experimentally measured value of
$\sim$~3.5 eV\cite{belomoin2000}, while the crystalline cluster with
reconstructed dimers are in agreement with experiment as previously
shown~\cite{mitas2001}.~ A comparison with oxygenated clusters
indicates that those with bridged oxygen are significantly higher than
our experimental gaps, while those with double bonded oxygen are
significantly lower, in both cases differing by over 1 eV.~ However,
these gaps are consistent with the calculated gaps of double core
amorphous-like clusters~\cite{draeger}, suggesting both structures
should be analyzed further.

\begin{table}[ht]
\begin{center}
\begin{tabular}{lcccc}
\smallskip
                        & ~~~LDA~~~& ~~~B3LYP~~~  & ~~~QMC~~~  \\
\hline
Si$_{29}$H$_{36}$       & 3.6     &  5.2        &  5.3(1)    \\
Si$_{35}$H$_{36}$       & 3.4     &  5.1        &  5.0(1)    \\
Si$_{29}$H$_{24}$       & 2.6     &  4.0        &  3.5(1)    \\
Si$_{30}$H$_{22}$       & 2.2     & (~~~~)      &  3.1(1)    \\
Si$_{29}$H$_{34}$O      & 3.1     &  4.8        &  4.7(1)    \\
Si$_{35}$H$_{34}$O      & 2.2     &  3.9        &  2.6(1)    \\
Si$_{35}$H$_{24}$O$_6$  & 1.7     &  3.3        & ( 1.7 )    \\
\end{tabular}
\caption
{Calculated optical gaps (eV) of some prototype 1~nm clusters using
  three different methods.  The statistical errors in the QMC values
  are in parenthesis.}
\end{center}
\end{table}

In order to better understand the possibility of ($2~\times~1$) dimer
reconstructions, we consider here the thermodynamics of the
reconstruction.  While the kinetics of such a reconstruction are
beyond the scope of this paper, to demonstrate that a ($2~\times~1$)
reconstruction is possible, we compared the total energies of the
``reaction" Si$_{29}$H$_{36} \,\to\,$ Si$_{29}$H$_{34} + H_2 $.~ We
found that within GGA methods, the balance is $\sim$ -0.3 eV
(endothermic at T=0) and therefore such a dimerization may well occur
for suitable chemical potentials of hydrogen.  Since the preparation
of the nanocrystals is performed in a mixture of HF and H$_2$O$_2$,
and indeed the presence of the peroxide is crucial for obtaining the
spherical shapes and nearly uniform sizes, one could also envisage the
reaction Si$_{29}$H$_{36} + $ H$_2$O$_2\,\to\,$ Si$_{29}$H$_{34}$ +
2H$_2$O.~ Since the oxygen-oxygen distance fits reasonably well with
the neighboring hydrogens on the two Si atoms such a reaction suggests
a short reaction path and the process is exothermic in GGA by
$\sim$~2.7~eV.~ Although the presence of H$_2$O$_2$ could also induce
additional oxidation reactions of the cluster, these reactions produce
clusters with absorption gaps considerably smaller than those measured
here when double bonded to the surface, and larger when in a bridged
configuration~\cite{puzder2002b}.~

Recently, DFT calculations with the B3LYP functional have received
some attention as a relatively computationally inexpensive alternative
to such many-body methods as couple cluster, GW-BSE, and QMC for
determining accurate absorption
gaps~\cite{garoufalis,williamson2002,zhou}.~ Table~1 shows calculated
gaps of Si$_{29}$H$_{34}$O with oxygen bridged to the surface, and
Si$_{35}$H$_{34}$O and Si$_{35}$H$_{24}$O$_{6}$ with oxygen double
bonded to the surface using the B3LYP functional.  In each case, the
gaps of the clusters with double bonded oxygen are significantly lower
than the observed gaps of our 1~nm clusters but in complete
disagreement with recent QMC studies.  Based on the B3LYP gaps, one
would conclude that our clusters are passivated by multiple double
bonded oxygen atoms, specifically Si$_{35}$H$_{24}$O$_{6}$ (3.3~eV
B3LYP gap), while the QMC results indicate otherwise.  Conversely, the
double core structures have gaps 1~eV higher than the experimentally
measured values, when calculated within B3LYP, which would tend to
eliminate these clusters as candidates.  Therefore, we conclude
through comparison of the absorption gap with QMC values, and through
simple thermodynamic considerations, that these clusters either have
($2\times1$) reconstructed surfaces, although they still may be
amorphous, and that like other DFT functionals, B3LYP may only
generate trends that are in agreement with QMC and not quantitative
values.

\section{Doping and Contamination of Reconstructed Silicon Clusters}

In the previous section, we calculated the effects of oxygen and
compared with our 1~nm prototype cluster with a reconstructed surface.
In this section, we consider the effect of other contaminants,
dopants, and functionalizing groups bonded to the surface of these
nanoclusters.  We have calculated the absorption gap of reconstructed
clusters with a variety of groups and once again compared the optical
gaps predicted by QMC with the single-particle B3LYP gaps.~ In our
previous study of {\it unreconstructed} clusters, double-bonded groups
were found to reduce the gap of 1~nm clusters by as much as $2.5$~eV,
while single-bonded groups reduced the gap a negligible
amount~\cite{puzder2002a}.~ Therefore, the completely hydrogenated
Si$_{29}$H$_{36}$ cluster yields a 5.3~eV gap, higher than our
observed 3.5~eV gap, while clusters with double-bonded passivants
yield gaps much smaller (2.0 to 2.7~eV).~ This supports the ($2 \times
1$) reconstructed Si$_{29}$H$_{24}$ cluster with a gap of 3.5~eV as a
likely candidate structure for our experiment~\cite{mitas2001}.~ To
complete the picture, we now consider the additional effect of
passivant groups on this reconstructed cluster, which has heretofore
{\it not} been considered.
  
Table 2 shows that our calculated LDA and B3LYP single-particle gaps
for a range of groups single-bonded to the surface of
Si$_{29}$H$_{24}$.~ We consider common contaminants from our synthesis
process (F and OH), groups typically used to dope semiconductors
(NH$_2$ and PH$_{2}$), and groups used to functionalize the surface
(CH$_{3}$ and SH).~ For all these groups, the reduction of the
single-particle gap compared to the prototype Si$_{29}$H$_{24}$
cluster is minimal ($<~0.1$~eV) similar to the small effect of
single-bonded groups on unreconstructed clusters~\cite{puzder2002a}.~
We find that this trend exists for calculations based on the LDA, PBE,
and B3LYP functionals.  This lack of an effect is perhaps not
surprising given that the bonding network between the surface and
these single bonded passivants is the same as with hydrogen.  The
interplay between the dimerization (the already distorted $sp^3$
network) and these various passivants is thus negligible demonstrating
that the dominant effect on the gap of Si$_{29}$H$_{24}$ is the
surface reconstruction, not the presence of single-bonded passivants.

\begin{table}[ht]
\begin{center}
\begin{tabular}{lcc}
\smallskip
Appendage Doping    & LDA/PBE & B3LYP \\
\hline
Si$_{29}$H$_{23}$CH$_{3}$               & 2.7   & 3.9      \\
Si$_{29}$H$_{23}$NH$_{2}$               & 2.6   & 3.9      \\
Si$_{29}$H$_{23}$SH                     & 2.5   & 3.7      \\
Si$_{29}$H$_{23}$C$_{4}$H$_{8}$SH       & 2.6   & 3.9      \\
Si$_{29}$H$_{23}$OH                     & 2.5   & 3.8      \\
Si$_{29}$H$_{23}$F                      & 2.6   & 3.8      \\
Si$_{29}$H$_{23}$PH$_{2}$               & 2.7   & 3.9      \\
\hline
Bridge Doping    \\
\hline 
Si$_{29}$H$_{24}$CH$_{2}$               & 2.7   & 4.0      \\
Si$_{29}$H$_{24}$NH                     & 2.6   & 3.8      \\ 
Si$_{29}$H$_{24}$S                      & 2.7   & 3.9      \\
\end{tabular}
\caption
{Calculated optical gaps (eV) of dopants on reconstructed clusters
  using various functionals within DFT either connected to one silicon atom
  (appendage doping) or between two atoms (bridge doping). }
\label{tab_dopes}
\end{center}
\end{table}

While the effect of single-bonded passivants is small, we found that
the SH and OH groups affect the nanocrystal gap more than other
single-bonded passivants yielding a 0.2~eV red shift, similar to the
observed gap reduction in OH on an unreconstructed
cluster~\cite{puzder2002a}.~ In each case, the addition of the bent
group tends to distort the binding geometry at the surface, be it
Si$_{35}$H$_{35}$OH, or Si$_{29}$H$_{23}$SH.~ The addition of a longer
hydrocarbon chain to the SH (C$_{4}$H$_{8}$SH), completely eliminates
this red shift as now the longer hydrocarbon chain causes less
distortion at the surface.  Therefore, our results show that caution
should be used when using single atoms to model the effect of foreign
substances on the surface of silicon nanoclusters; these single atom
models tend to distort the surface to a greater extent and thus
overestimate the affect on the gap compared with longer chains.

Dopants may also form in a bridged configuration which may potentially
lower the gap further when coupled with dimerization.  Previous
calculations have shown that in Si$_{29}$ clusters, the formation of
bridged oxygen is energetically favorable to double-bonded
oxygen~\cite{puzder2002b}.~ In Table~2, we compare the gaps of a
number of additional dopants in an Si-X-Si configuration with X = S,
NH, and CH$_{2}$.~ The Si-X-Si bridge replaces a reconstructed dimer
on the surface.  Again, we find the effect of these bridged dopants is
negligible.

These minimal shifts in the optical absorption gap mean that
single-bonded and bridged contaminants cannot be distinguished from
fully hydrogenated clusters in optical absorption measurements and
thus may be present on our clusters.  The calculation of other
physical properties such as the Stokes shift are then required for
additional characterization.

\section{Stokes Shift}

\subsection{Mechanism}

One of the most intriguing features of nanocrystals is the possibility
of forming self-trapped excitons which are closely related to the
Stokes shift.  This possibility has been the subject of several recent
studies using a variety of
models~\cite{martin,takagahara,allan96,puzder2003}.~ In particular,
Allan~{\em et al.}  have examined models involving both a relaxation
mechanism involving the entire cluster~\cite{martin} and more recently
a model based on the assumption that, after absorption, the exciton
leads to a stretching of a particular surface Si-Si dimer bond, to
form a self-trapped exciton~\cite{allan96}.~ The motivation for
reexamining a range of different models arises from the observation
that many experiments on 1~nm clusters appear to lead to pronounced
Stokes shifts. However, in our experiments we obtained values of
$\sim$ 0.4 - 0.5 eV~\cite{belomoin2000} which are {\em significantly
  smaller} than previously calculated $\sim$ 1-2 eV for this range of
sizes~\cite{allan96,wolkin}.~ By calculating the Stokes shift in ideal
1~nm clusters, in those with reconstructed surfaces, in those with
bridged oxygen, and in double-core amorphous-like clusters, we hope to
demonstrate the use of the Stokes shift as a useful characterization
method.
  
Before comparisons of the excitations in different structural
prototypes can be made, we first need to determine the appropriate
relaxation mechanisms.  To resolve this issue we have performed {\em
  ab initio} calculations of the relaxation of the candidate
structures discussed above in both the ground and excited (excitonic)
electronic states.  In addition, we have used QMC calculations to
provide highly accurate values for the electronic gaps of the ground
and excited state structures to accurately determine the magnitude of
the Stokes shift associated with each model.  Since it is
computationally easier to relax the electronic structure in the
optically forbidden triplet excited state, we used the triplet state
to carry out most of geometry scanning and relaxation calculations.
For a few points on energy surfaces we have verified that the singlet
and triplet energies were within $\sim$ 0.03 eV.~ Therefore, the
measured Stokes shift is about an order of magnitude larger than the
singlet-triplet splitting, in direct contradiction to the model
proposed by Takagahara~\cite{takagahara} for the Stokes shift in
silicon nanoclusters.
  
\begin{figure}
\rotatebox{-90}{
\scalebox{0.3}{
\includegraphics{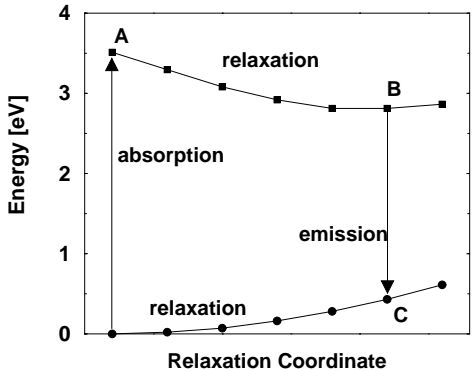}
}
}
\caption{Illustration of the Stokes shift in Si$_{35}$H$_{34}$ within the 
  collective relaxation mechanism using energies from local density
  approximation calculations.}
\label{fig:glob}
\end{figure}

To compare different Stokes shift mechanisms, we choose to study the
Si$_{29}$H$_{34}$ structure.  This cluster has the same structure as
the Si$_{29}$H$_{36}$ prototype, except that one of the -SiH$_2$
surface pairs has been reconstructed to form a dimer.  By studying the
Stokes shift in this cluster we are able to examine the competition
between a dimer localized Stokes shift and a global relaxation of the
cluster.  In addition, we are able to compare with previous
studies~\cite{allan96,hirao,caldas2000} of this system.  In
Fig.~\ref{fig:glob} we plot the LDA total energy of a
Si$_{29}$H$_{34}$ nanocrystal in both the ground state (lower curve:
circles) and excited state using the ground state structure total
energy as a reference.  Once the photon is absorbed (point A) the
system is in the excited state and the slower ($\approx$ picoseconds)
structural relaxation occurs.

The decrease in the energy from the point A to B is due to the
collective relaxation of all atomic positions of the entire cluster
without any constraint except that the system is in the excited
electronic state.  In the two symmetric clusters, either the
unreconstructed Si$_{29}$H$_{36}$, or the Si$_{29}$H$_{24}$, the
excitation is formed by promoting an electron from the $p$-like HOMO
to the $s$-like LUMO, leading to a small distortion of the cluster
from a spherical to slightly elliptical geometry. The position C
corresponds to the state after photon emission and before the
subsequent relaxation to the ground state.  The key features which
emerge from our calculations are: a) barrierless relaxation to a lower
energy geometry and b) small geometry adjustments of $\sim$ 0.01 A of
essentially all atoms in the cluster.

\begin{figure}
\rotatebox{-90}{
\scalebox{0.3}{
\includegraphics{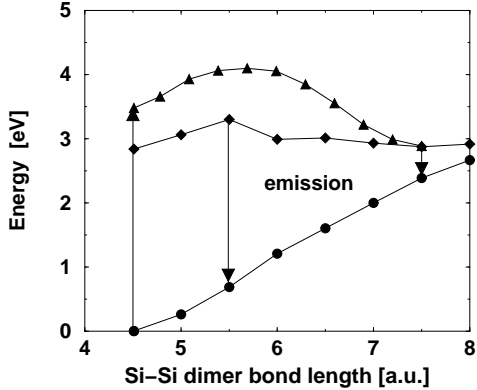}
}
}
\caption{Density functional Stokes shift calculations of Si$_{35}$H$_{34}$ 
within the local bond-breaking relaxation mechanism.}
\label{fig:loc}
\end{figure}

In Fig.~\ref{fig:loc} we analyzed the model of a self-trapped exciton
based on a stretching and breaking of the single dimer bond, as
proposed in Ref.~\cite{allan96}.  The geometries used to generate the
upper triplet curve (triangles) were obtained by linearly
interpolating the atomic positions between the ground state geometry
($d_{Si-Si} =2.4$ \AA) and the local minimum energy structure obtained
when $d_{Si-Si}$ was constrained to 4.0 \AA.  This upper curve closely
reproduces the calculation originally presented in Fig.~2 of
Ref.~\cite{allan96} for this system, where the intermediate structures
were also derived from linear interpolation.  The points on the lower
curve (diamonds) were obtained by constraining the dimer bond length,
$d$, to a series of different values, while allowing all other atoms
to relax while keeping the system in the excited state.

Examining Fig.\ref{fig:loc}, we observe that if the system goes along
the structural path leading to the broken dimer bond, it first has to
overcome a barrier of $\sim$ 0.5 eV due to the elastic energy of the
cluster before the energy decreases as the bond breaks
($d_{Si-Si}>2.9$ \AA).  It is interesting that the minimum energy of
the excited state resulting from the global relaxation in
Fig.\ref{fig:loc} is almost identical (2.8 eV) to the minimum obtained
by breaking the dimer, 2.9 eV. However, the important difference is in
the large energy increase on the ground state path which in turn leads
to a very large Stokes shift and small energy of the emitted photon.
The curve represented by interpolating the atomic coordinates between
the ground state structure and the structure of a broken dimer
($d_{Si-Si} \sim $ 4.0 \AA) as suggested in Ref.~\cite{allan96} yields
a markedly different energy surface to the more realistic case in
which all the atoms, except the dimer, are relaxed for each dimer
length.

We have also examined the Stokes shift relaxation mechanism proposed
in Ref.~\cite{caldas2000} where a hydrogen atom attached to one of two
neighboring Si-H surface groups moves into a bridged position between
the silicon atoms and the Si-Si bond stretches.  In contrast to the
semi-empirical calculations used to predict this
structure~\cite{caldas2000}, our density functional calculations do
not find this structure to be a meta-stable state.  We find a
spontaneous relaxation from this proposed bridge structure to a
structure in which the bridged hydrogen is completely transfered to
the neighboring Si, producing an Si-H$_2$ group and a Si with a
dangling bond.  Interestingly, this structure is energetically
competitive (3.1 eV above groundstate) with the relaxed excited state
structures shown in Figs.~\ref{fig:glob} and \ref{fig:loc}.  However,
as with the dimer breaking mechanism illustrated in
Fig.~\ref{fig:loc}, there is a significant energy barrier between the
groundstate structure and this one.
  
Based on these results, three different mechanisms from which a Stokes
shift could result are possible.  After absorbing a photon the cluster
could then: i) relax via the collective structural mechanism
(Fig.~\ref{fig:glob}) to point (B) where the electron and hole then
vertically recombine to point (C); or, ii) absorb enough energy from
thermal fluctuations or higher vibronic states~\cite{nayfeh97} to
overcome the barrier and either break the dimer~\cite{allan96} or
transfer a hydrogen~\cite{caldas2000} with subsequent emission and
relaxation to the ground state (Fig.~\ref{fig:loc}, triangles); or,
iii) absorb enough energy from thermal excitations to partially
stretch the dimer and to recombine from the top of the barrier.  The
LDA values of the Stokes shift for these three mechanisms are 1.1, 3.0
and 0.9 eV, respectively.  The significantly larger Stokes shift for
mechanism ii) arises mostly from the large increase in the ground
state energy associated with the stretching of the dimer bond.  At the
minimum of the excited state, the dimer is effectively broken and the
ground state energy has increased by $\sim$ 2.5 eV, the energy
required to create two dangling bonds.

Given the very large Stokes shift associated with the creation of a
surface dimer or transfer of hydrogen, coupled with the significant
barrier that first has to be overcome before it is energetically
favorable to stretch the dimer, we believe that the mechanism most
likely to be responsible for the Stokes shift is the global relaxation
mechanism of Fig.~\ref{fig:loc}.  We have therefore calculated within
both DFT (LDA and GGA) and QMC the values of the total energies at
points A to C for the three candidate structures.  

\subsection{Results}

The QMC values for the Stokes shifts (Table~3), defined as
$(E_A-E_{\rm ground})-(E_B-E_C)$, for Si$_{29}$H$_{36}$,
Si$_{29}$H$_{24}$ and Si$_{35}$H$_{36}$ are 1.1, 0.42 and 0.8 eV,
respectively, and agree well with those predicted by LDA and PBE which
are about 0.1~eV higher in each case.  While the decrease in value 
between the Si$_{29}$H$_{36}$ and the Si$_{35}$H$_{36}$ cluster demonstrates 
clearly the size dependence of ideal spherical silicon nanoclusters, the
QMC value of Si$_{29}$H$_{24}$ is in closest agreement with the
measured value of $\sim $ 0.45 eV~\cite{belomoin2002}.~ Therefore,
when combined with our estimation for the gap, 
Si$_{29}$H$_{24}$ remains a realistic prototype for both the structure and
excited states processes observed in our experiments.

\begin{table}[ht]
\begin{center}
\begin{tabular}{lcc}
\smallskip         
                        & \multicolumn{2}{c}{Stokes Shift}     \\
                        & ~~~~LDA~~~~       &  ~~~~QMC~~~~  \\
\hline
Si$_{29}$H$_{36}$       & 1.1               &  1.0        \\
Si$_{35}$H$_{36}$       & 0.8               &  0.7        \\
Si$_{29}$H$_{24}$       & 0.5               &  0.45       \\
Si$_{30}$H$_{22}$       & 3.0               &             \\
\end{tabular}
\caption
{Calculated Stokes shifts (eV) of some prototype 1~nm clusters 
  calculated within LDA and QMC.  The large shift in Si$_{30}$H$_{22}$
  renders the QMC unnecessary.  }
\end{center}
\end{table}

We also consider here the non-crystalline 1~nm clusters which have been 
predicted to form during chemical vapor deposition at high temperatures by         
quantum molecular dynamics (QMD) simulations~\cite{draeger}.~  Although not applicable 
to the sonification process demonstrated here, these non-crystalline clusters 
were shown to have a gap comparable to those with ideal reconstructed clusters.  
We calculate the Stokes shift of these clusters 
to ascertain whether they have the 0.4 to 0.5 eV Stokes shifts observed in our
clusters and predicted for reconstructed clusters.  
Surprisingly, we find very large Stokes shifts, on the order of 
the gap size!  These non-crystalline 1~nm clusters behave more like small molecules~\cite{franceschetti} 
than quantum dots.  Thus, the Stokes 
shift has proved a very powerful characterization technique 
eliminating the ``double core'' non-crystalline clusters as being those
observed here.

\section{Conclusion}

In conclusion, we have carried out a thorough study of hydrogen
terminated silicon nanoclusters in the 1 - 1.2 nm range.  We
investigated several prototype structures and compared their optical
absorption gaps and Stokes shifts with recent measurements and
found that although a few other structures may yield a similar gap, most 
notably an amorphous-like double core cluster, only Si$_{29}$H$_{24}$ 
yields both the correct gap and Stokes shift.  Thus,
we determine that {\em both} properties must be considered when
evaluating candidate structures to interpret optical measurements.
We determined that B3LYP generates inconsistent results for clusters 
with localized orbitals compared with QMC level calculations.  The 
atomistic first-principles DFT approaches coupled with QMC allowed us to
study the optically induced excitons and to conclude that the most
likely mechanism causing the Stokes shift is the barrierless
relaxation of the whole structure with the red shift of $\sim$~0.4~eV
in agreement with experiment.  Comparison with further experimental
data indicates that the Si$_{29}$H$_{24}$ structural prototype is the
most promising candidate of the possibilities we tested.  We have also
investigated the effect of doping with a number of atoms and molecular
groups. Like ideal unreconstructed structures, $sp^3$ bonded passivants 
have a minimal effect on absorption gaps, as do single bridged passivants.
Therefore, other atoms or ligands may be used to functionalize these 
clusters with no discernible change to the gap.

We thank Erik Draeger for providing QMD non-crystalline structures and for helpful discussions. 
In part this work was performed under the auspices of the U.S. Department of Energy
by the University of California, Lawrence Livermore National
Laboratory under contract No. W-7405-Eng-48.  We also gratefully
acknowledge NSF support from the grant DMR-0102668 and M.N.
acknowledges support by the NSF grant. Part of the calculations has
been done at NCSA, University of Illinois.



  

\end{document}